\documentclass[10pt,conference,letterpaper]{IEEEtran}
\IEEEoverridecommandlockouts

\usepackage{amssymb,amsmath,amsthm, mathtools}
\usepackage{graphicx}
\usepackage{cite}
\usepackage{latexsym}
\usepackage{verbatim}
\usepackage{amsmath}
\usepackage{amssymb}
\usepackage{subfigure}
\usepackage{url}
\usepackage{epstopdf}
\usepackage{algorithm}
\usepackage{times}
\usepackage{stfloats}
\usepackage{multirow}
\usepackage{enumerate}
\usepackage{xspace}
\usepackage{longtable}
\usepackage{mathrsfs}
\usepackage{mathtools}
\usepackage{algcompatible}
\usepackage[noend]{algpseudocode}
\usepackage{amssymb}
\usepackage{pifont}
\usepackage{multicol}
\usepackage{caption}
\usepackage{color}
\usepackage{tikz}

\makeatletter
\newcommand{\algrule}[1][.2pt]{\par\vskip.5\baselineskip\hrule height #1\par\vskip.5\baselineskip}
\makeatother
\newcommand{\PPSS}{\ensuremath {\mathit{PPSS}}{\xspace}}
\newcommand{\fc}{\ensuremath {\mathit{FC}}{\xspace}}

\newcommand{\crn}{\ensuremath {\mathit{CRN}}{\xspace}}

\newcommand{\su}{\ensuremath {\mathit{SU}}{\xspace}}
\newcommand{\decision}{\ensuremath {\mathit{dec}}{\xspace}}
\newcommand{\loDec}{\ensuremath {\mathit{D}}{\xspace}}
\newcommand{\loHyp}{\ensuremath {\mathit{h}}{\xspace}}
\newcommand{\Hyp}{\ensuremath \mathcal{H}{\xspace}}
\newcommand{\pu}{\ensuremath {\mathit{PU}}{\xspace}}
\newcommand{\gw}{\ensuremath {\mathit{GW}}{\xspace}}
\newcommand{\rss}{\ensuremath {\mathit{RSS}}{\xspace}}
\newcommand{\ta}{\ensuremath {\mathit{\tau}}{\xspace}}
\newcommand{\lam}{\ensuremath {\mathit{\lambda}}{\xspace}}

\newcommand{\ope}{\ensuremath {\mathit{OPE}}{\xspace}}
\newcommand{\key}{\ensuremath {\mathit{k}}{\xspace}}
\newcommand{\ciphFC}{\ensuremath {\mathit{c}}{\xspace}}
\newcommand{\ciphU}{\ensuremath {\mathit{\varsigma}}{\xspace}}
\newcommand{\rsa}{\ensuremath {\mathit{N}}{\xspace}}
\newcommand{\ciphGW}{\ensuremath {\mathit{\zeta}}{\xspace}}
\newcommand{\user}{\ensuremath {\mathit{U}}{\xspace}}
\newcommand{\nbr}{\ensuremath {\mathit{n}}{\xspace}} 
\newcommand{\gam}{\ensuremath {\mathit{\gamma}}{\xspace}} 
\newcommand{\bin}{\ensuremath {\mathit{b}}{\xspace}} 
\newcommand{\votes}{\ensuremath {\mathit{v}}{\xspace}} 

\newcommand{\sr}{\ensuremath {\mathit{\mu}}{\xspace}}
\newcommand{\chr}{\ensuremath {\mathit{\beta}}{\xspace}}

\newcommand{\kap}{\ensuremath {\mathit{\kappa}}{\xspace}}\newcommand{\posr}{\ensuremath {\mathit{\varrho}}{\xspace}}
\newcommand{\ran}{\ensuremath {\mathit{R}}{\xspace}}
\newcommand{\alhv}{\ensuremath {\mathit{\alpha}}{\xspace}} 

\newcommand{\ym}{\ensuremath {\mathit{YM}}{\xspace}}

\newcommand{\aes}{\ensuremath {\mathit{AES}}{\xspace}} 
\newcommand{\blck}{\ensuremath {\mathit{\epsilon_{\En}}}{\xspace}} 
\newcommand{\pr}{\ensuremath {\mathit{\pi}}{\xspace}}

\newcommand{\OEnc}[2]{\ensuremath{\mathit{OPE}.\mathcal{E}_{#1}\mskip-1mu(#2)}}
\newcommand{\OEncc}{\ensuremath{\mathit{OPE}.\mathcal{E}}}
\newcommand{\LU}{\ensuremath{\mathcal{L}_1}}
\newcommand{\LF}{\ensuremath{\mathcal{L}_2}}
\newcommand{\LG}{\ensuremath{\mathcal{L}_3}}
\newcommand{\V}{\ensuremath{\vec{V}}}

\newcommand{\A}{\ensuremath{\mathcal{A}}}

\newcommand{\as}{\ensuremath {\leftarrow}{\xspace}}

\newcommand{\En}{\ensuremath{\mathcal{E}}{\xspace}}
\newcommand{\De}{\ensuremath{\mathcal{D}}{\xspace}}
\newcommand{\Enc}[2]{\ensuremath{\mathcal{E}_{#1}\mskip-1mu(#2)}}
\newcommand{\Dec}[2]{\ensuremath{\mathcal{D}_{#1}\mskip-1mu(#2)}}

\newcommand{\LPOS}{\ensuremath {\mathit{LPOS}}{\xspace}}

\newcommand{\PDAFT}{\ensuremath {\mathit{PDAFT}}{\xspace}}
\newcommand{\LPGW}{{\em LP-3PSS}{\xspace}}   

\newcommand{\srv}{\ensuremath {\mathit{y}}{\xspace}}
\makeatother

{\bfseries}{\rmfamily}
\newtheorem{definition}{Definition}{\bfseries}{\rmfamily}
\newtheorem{mytheorem}{Theorem}{\bfseries}{\rmfamily}
{\bfseries}{\rmfamily}
\newtheorem{mycorollary}{Corollary}{\bfseries}{\rmfamily}
\newtheorem{fact}{Fact}{\bfseries}{\rmfamily}
\newtheorem{assumption}{Security Assumption}{\bfseries}{\rmfamily}
\newtheorem{objective}{Security Objective}{\bfseries}{\rmfamily}
{\bfseries}{\rmfamily}

\newcommand\copyrighttext{%
  \footnotesize \copyright2016 IEEE. Personal use of this material is permitted. Permission from IEEE must be obtained for all other uses, in any current or future media, including reprinting/republishing this material for advertising or promotional purposes, creating new collective works, for resale or redistribution to servers or lists, or reuse of any copyrighted component of this work in other works.}

\begin{document}

  \title{An Efficient Technique for Protecting Location Privacy of Cooperative Spectrum Sensing Users
  \vspace{-0.2in}
}

\author{Mohamed~Grissa$^{\star}$, Attila Yavuz$^{\star}$, and Bechir Hamdaoui$^{\star}$\\
$^{\star}$\small Oregon State University, grissam,yavuza,hamdaoub@onid.oregonstate.edu\\
\vspace{-5.5mm}
\thanks{\copyrighttext}
}

\maketitle
{\let\thefootnote\relax\footnote{{\\Digital Object Identifier 10.1109/INFCOMW.2016.7562209}}}

\begin{abstract}
Cooperative spectrum sensing, despite its effectiveness in enabling dynamic spectrum access, suffers from location privacy threats, merely because secondary users (\su s)' sensing reports that need to be shared with a fusion center to make spectrum availability decisions are highly correlated to the users' locations. It is therefore important that cooperative spectrum sensing schemes be empowered with privacy preserving capabilities so as to provide \su s with incentives for participating in the sensing task. In this paper, we propose an efficient privacy preserving protocol that uses an additional architectural entity and makes use of various cryptographic mechanisms to preserve the location privacy of \su s while performing reliable and efficient spectrum sensing. We show that not only is our proposed scheme secure and more efficient than existing alternatives, but also achieves fault tolerance and is robust against sporadic network topological changes.

\end{abstract}


\section{Introduction}
\label{sec:Introduction}
{\em Cooperative spectrum sensing} is a key component of cognitive radio networks ($\crn$s) essential for enabling dynamic and opportunistic spectrum access~\cite{akyildiz2011}. It consists of having secondary users (\su s) sense the licensed channels on a regular basis and collaboratively decide whether a channel is available prior to using it so as to avoid harming primary users (\pu s).  %
One of the most popular spectrum sensing techniques is energy detection, thanks to its simplicity and ease of implementation, which essentially detects the presence of \pu~signal by measuring and relying on the energy strength of the sensed signal, commonly known as the received signal strength (\rss)~\cite{fatemieh2011using}.

Broadly speaking, cooperative spectrum sensing techniques can be classified into two categories: Centralized and distributed~\cite{akyildiz2011}. In centralized techniques, a central entity called fusion center (\fc) orchestrates the sensing operations as follows. It selects one channel for sensing and, through a control channel, requests that each \su~perform local sensing on that channel and send its sensing report (e.g., the observed \rss~value) back it to. It then combines the received sensing reports, makes a decision about the channel availability, and diffuses the decision back to the \su s. In distributed sensing techniques, \su s do not rely on a \fc~for making channel availability decisions. They instead exchange sensing information among one another to come to a unified decision~\cite{akyildiz2011}. This requirement makes distributed sensing techniques highly complex with respect to their centralized counterparts. Hence, centralized sensing techniques are considered more practical for real-life applications.

Despite its usefulness and effectiveness in promoting dynamic spectrum access, cooperative spectrum sensing suffers from serious security and privacy threats. One big threat to \su s, which we tackle in this work, is location privacy, which can easily be leaked due to the wireless nature of the signals communicated by \su s during the cooperative sensing process. In fact, it has been shown that \rss~values of $\su$s are highly correlated to their physical locations~\cite{li2012location}, thus making it easy to compromise the location privacy of \su s when sending out their sensing reports.
The fine-grained location, when combined with other publicly available information, could easily be exploited to infer private information about users~\cite{wicker2012loss}. Examples of such private information are shopping patterns, user preferences, and user beliefs, just to name a few~\cite{wicker2012loss}. With such privacy threats and concerns, $\su$s may refuse to participate in the cooperative sensing tasks.
It is therefore imperative that cooperative sensing schemes be enabled with privacy preserving capabilities that protect the location privacy of \su s, thereby encouraging them to participate in such a key $\crn$ function, the spectrum sensing.

In this paper, we propose an efficient privacy-preserving scheme for cooperative spectrum sensing that exploits various cryptographic mechanisms to preserve the location privacy of \su s while performing the cooperative sensing task reliably and efficiently. We show that our proposed scheme is secure and more efficient than its existing counterparts, and is robust against sporadic topological changes and network dynamism.

\subsection{Related Work}
\label{sec:related}
Security and privacy in $\crn$s have gained some attention recently. Yan et al.~\cite{6195839} discussed security issues in fully distributed cooperative sensing. Qin et al.\cite{qin2014preserving} proposed a privacy-preserving protocol for \crn~transactions using a commitment scheme and zero-knowledge proof.

Location privacy, though well studied in the context of location-based services (LBS)~\cite{6567111,6567112}, has received little attention in the context of \crn s~\cite{gao2013location,liu2013location,li2012location}. Some works focused on location privacy but not in the context of cooperative spectrum sensing (e.g., database-driven spectrum sensing~\cite{gao2013location,grissa2015cuckoo}  and dynamic spectrum auction~\cite{liu2013location}) and are skipped since they are not within this paper's scope.

In the context of cooperative spectrum sensing, Shuai et al.~\cite{li2012location} showed that \su s' locations can easily be inferred from their \rss~reports, and called this the SRLP (single report location privacy) attack. They also identified the DLP (differential location privacy) attack, where a malicious entity can estimate the \rss~(and hence the location) of a leaving/joining user from the variations in the final aggregated \rss~measurements before and after user's joining/leaving of the network. They finally proposed \PPSS, a protocol for cooperative spectrum sensing, to address these two attacks. Despite its merits, \PPSS~has several limitations: (i) It needs to collect all the sensing reports to decode the aggregated result. This is not fault tolerant, since some reports may be missing due, for e.g., to the unreliable nature of wireless channels. (ii) It cannot handle dynamism if multiple users join or leave the network simultaneously. (iii) The pairwise secret sharing requirement incurs extra communication overhead and delay. (iv) The underlying encryption scheme requires solving the {\em Discrete Logarithm Problem}\cite{mccurley1990discrete}, which is possible only for very small plaintext space and can be extremely costly (see Table~\ref{tab:Table2}).  Chen et al.~\cite{chen2014pdaft} proposed \PDAFT, a fault-tolerant and privacy-preserving data aggregation scheme for smart grid communications. \PDAFT, though proposed in the context of smart grids, is suitable for cooperative sensing schemes. But unlike \PPSS, \PDAFT~relies on an additional semi-trusted entity, called gateway, and like other aggregation based methods, is prone to the DLP attack. In our previous work \cite{grissa2015location} we proposed an efficient scheme called \LPOS~to overcome the limitations that existent approaches suffer from. \LPOS~combines order preserving encryption and yao's millionaire protocol to provide a high location privacy while enabling an efficient sensing performance.

\subsection{Our Contribution} \label{subsec:OurContribution}
In this paper, we propose a new location privacy-preserving scheme that we call \LPGW~(location privacy for 3-party spectrum sensing architecture), which harnesses various cryptographic primitives (e.g., order preserving encryption) in innovative ways along with an additional  architectural entity (i.e., a gateway) to achieve high location privacy with a low overhead. That is, our proposed \LPGW~scheme offers the following desirable properties:
\begin{itemize}
\item Location privacy of secondary users while performing the cooperative spectrum sensing effectively and reliably.

\item Fault tolerance and robustness against network dynamism (e.g., multiple \su s join/leave the network) and failures (e.g., missed sensing reports).

\item Reliability and resiliency against malicious users via an efficient reputation mechanism.

\item Accurate spectrum availability decisions via half-voting rules while incurring minimum communication and computation overhead.
\end{itemize}

Note that for simplicity we use energy detection through \rss~measurement for spectrum sensing in our scheme. However, our scheme can be applied with any other spectrum detection technique whose sensing reports may leak information about the location of the users.

\section{Preliminaries}
\label{sec:Preliminaries}
We consider a cooperative spectrum sensing architecture that consists of a \fc~and a set of \su s, where each \su~is assumed to be capable of measuring \rss~on any channel by means of an energy detection method~\cite{fatemieh2011using}.
In this cooperative sensing architecture, the \fc~combines the sensing observations collected from the \su s, decides about the spectrum availability, and broadcasts the decision back to the \su s through a control channel.
This could typically be done via either {\em hard} or {\em soft} decision rules. The most common soft decision rule is aggregation, where \fc~collects the \rss~values from the \su s and compares their average to a predefined threshold, \ta, to decide on the channel availability.

In hard decision rules, e.g. voting, \fc~combines votes instead of \rss~values. Here, each \su~compares its \rss~value with \ta, makes a local decision (available or not), and then sends to the \fc~its one-bit local decision/vote instead of sending its \rss~value. \fc~applies then a voting rule on the collected votes to make a channel availability decision.
However, for security reasons to be discussed shortly, it may not be desirable to share \ta~with \su s. In this case, \fc~can instead collect the \rss~values from the \su s, make a vote for each \su~separately, and then combine all votes to decide about the availability of the channel.

In this work, we opted for the voting-based decision rule, with \ta~is not to be shared with the \su s, over the aggregation-based rule. There are two reasons for choosing voting-based decision rule over the aggregation-based decision rule: (i) Aggregation methods are more prone to sensing errors; for example, receiving some erroneous measurements that are far off from the average of the \rss~values can skew the computed \rss~average, thus leading to wrong decision. (ii) Voting does not expose users to the DLP attack~\cite{li2012location} (which is identified  earlier in Section~\ref{sec:related}).
We chose not to share \ta~with the \su s because doing so limits the action scope of malicious users that may want to report falsified \rss~values for malicious and/or selfish purposes.

In this paper we investigate a 3-party cooperative sensing architecture, where a third entity, called gateway (\gw), is incorporated along with the \fc~and \su s to cooperate with them in performing the sensing task. As will be shown later, this additional gateway allows to achieve higher privacy and lesser computational overhead, but of course at its cost.

\subsection{Security Threat Model and Objectives}
We consider a {\em semi-honest} threat model, where all the network parties (i.e., \su s, \fc, and \gw) are assumed to be {\em honest but curious} in that they execute the protocol honestly but show interest in learning information about the other parties. This means that none of these entities is trusted. More specifically, we make the following assumptions:
\begin{assumption} \label{asm:asm1}
No party in the system modifies maliciously (or nonmaliciously) the integrity of its input. That is, (i) \fc~does not maliciously inject false \ta; and (ii) the \su s do not maliciously change their \rss~values.
\end{assumption}
\begin{assumption} \label{asm:asm2}
No party in the system colludes with any of the other parties. That is, (i) \fc~does not collude with \su s; (ii) \su s do not collude with one another; and (iii) \gw~does not collude with \su s or \fc.
\end{assumption}

As mentioned before, \rss~values are shown to be highly correlated to the \su s' locations~\cite{li2012location}. Therefore, if the confidentiality of the \rss~values is not protected, then nor is the location privacy of the \su s.
With this in mind, the security objectives of the proposed schemes are then:
\begin{objective} \label{obj:SecurityObj-1}
Keep the \rss~value of each \su~confidential to the \su~only by hiding it from all other parties. This should hold during all sensing periods and for any network membership change.
\end{objective}

Also, since \su s may rely on the threshold \ta~to maliciously manipulate their \rss s, our second objective is then to:
\begin{objective} \label{obj:SecurityObj-2}
Keep \ta~confidential to the \fc~only by hiding it from all other parties. This should hold during all sensing periods and for any network membership change.
\end{objective}

\subsection{Half-Voting Availability Decision Rule}
\label{subsec:half-voting}
Our proposed scheme uses the {\em half-voting decision rule}, shown to be optimal in~\cite{zhang2008cooperative}, and for completeness, we here highlight its main idea.
Details can be found in~\cite{zhang2008cooperative}.

Let $\loHyp_0$ and $\loHyp_1$ be the spectrum sensing hypothesis that \pu~is absent and present, respectively. Let $P_f$, $P_d$ and $P_m$ denote the probabilities of false alarm, detection, and missed detection, respectively, of one \su; i.e., $P_f = Pr(\rss>\ta\mid \loHyp_0)$, $P_d = Pr(\rss>\ta\mid \loHyp_1)$, and $P_m = 1 - P_{d}$.
%
%
\fc~collects the 1-bit decision $\loDec_i$ from each \su~$U_i$ and fuses them together according to the following fusion rule~\cite{zhang2008cooperative}:
\begin{equation} \label{FCHyp}
\decision =  \begin{cases}  \Hyp_1, & \displaystyle\sum \limits _{i=1}^n \loDec_i\geq \lam \\  \Hyp_0, & \displaystyle\sum \limits _{i=1}^n \loDec_i <\lam\end{cases}
\end{equation}
Note that \fc~infers that \pu~is present when at least \lam~\su s are inferring $\loHyp_1$. Otherwise, \fc~decides that \pu~is absent, i.e. $\Hyp_0$. Note here that the OR fusion rule corresponds to the case where $\lam=1$ and the AND fusion rule corresponds to the case where $\lam=\nbr$.
The cooperative spectrum sensing false alarm probability, $Q_f$, and missed detection probability, $Q_m$, are: $Q_f = Pr(\Hyp_1\mid\loHyp_0)$ and $Q_m = Pr(\Hyp_0\mid\loHyp_1)$.
Letting $n$ be the number of \su s, the optimal value of \lam~that minimizes $Q_f+Q_m$ is $\lam_{opt} = \min(\nbr, \lceil{\nbr}/{(1+\alhv)}\rceil)$,
%
where $\alhv = \ln (\frac{P_f}{1-P_m})/\ln (\frac{P_m}{1-P_f})$ and $\lceil\cdot\rceil$ denotes the ceiling function.
For simplicity, $\lam_{opt}$ is denoted as $\lam$ throughout this paper.

\subsection{Reputation Mechanism}

To make the voting rule more reliable, we incorporate a reputation mechanism that allows \fc~to progressively eliminate faulty and malicious \su s. It does so by updating and maintaining a reputation score for each \su~to reflect the level of reliability the \su~has. Our proposed schemes incorporate the {\em Beta Reputation} mechanism, proposed and shown to be robust by Arshad et al.~\cite{arshad2011robust}. For completeness, we highlight its key features next; more details can be found in~\cite{arshad2011robust}.

At the end of each sensing period $t$, \fc~obtains a decision vector, $\mbox{\boldmath$ \bin$}(t)=[\bin_1(t),\bin_2(t),\ldots,\bin_\nbr(t)]^T$ with $\bin_i(t) \in \{0,1\}$, where $\bin_i(t)=0$ (resp. $\bin_i(t)=1$) means that the spectrum is reported to be free (resp. busy) by \su~$U_i$. \fc~then makes a global decision using the fusion rule $f$ as follows:
\begin{equation}
\label{fDef}
\decision(t)=f(\boldsymbol{w}(t),\mbox{\boldmath$ \bin$}(t)) = \begin{cases} 1 & \mbox{if } \sum\limits_{i=1}^{\nbr}w_i(t) \bin_i(t) \geq\lam \\ 0 & \mbox{otherwise }\end{cases}
\end{equation}
where $\boldsymbol{w}(t) = [w_1(t),w_2(t)\ldots,w_\nbr(t)]^T$ is the weight vector calculated by \fc~based on the credibility score of each user, as will be shown shortly, and $\lam$ is the voting threshold determined by the Half-voting rule~\cite{zhang2008cooperative}, as presented in Section~\ref{subsec:half-voting}.

For each \su~$U_i$, \fc~maintains positive and negative rating coefficients, $\posr_i(t)$ and $\eta_i(t)$, that are updated every sensing period $t$ as: $\posr_i(t) = \posr_i(t - 1) +\nu_1(t)$ and $\eta_i(t) = \eta_i(t - 1) +\nu_2(t)$, where
$\nu_1(t)$ and $\nu_2(t)$ are calculated as
\vspace{-20pt}
\begin{multicols}{2}
\begin{equation*}
\nu_1(t) = \begin{cases} 1 & \bin_i(t) = \decision(t)\\ 0 & \mbox{otherwise }\end{cases}
\end{equation*}\break
\begin{equation*}
\nu_2(t) = \begin{cases} 1 & \bin_i(t) \neq \decision(t)\\ 0 & \mbox{otherwise }\end{cases}
\end{equation*}
\end{multicols}
Here, $\posr_i(t)$ (resp. $\eta_i(t)$) reflects the number of times $U_i$'s observation, $\bin_i(t)$, agrees (resp. disagrees) with the $\fc$'s global decision, $\decision$(t).

\fc~computes then $U_i$'s credibility score, $\varphi_i$(t), and contribution weight, $w_i$(t), at sensing period $t$ as:
\vspace{-30pt}
\begin{multicols}{2}
\begin{equation}
\label{cred}
\varphi_i (t) \!=\! \dfrac{\posr_i(t) + 1}{\posr_i(t) \! + \! \eta_i(t)\! +\! 2}
\end{equation}\break
\begin{equation}
\label{weight}
w_i (t)\!= \! {\varphi_i(t)}/{\sum\limits_{j=1}^{\nbr}\!\varphi_j(t)}
\end{equation}
\end{multicols}

\subsection{Cryptographic Building Blocks}
\label{subsec:crypto}
Our scheme uses a well known cryptographic building block, which we define next before using it in the next section when describing our scheme so as to ease the presentation.



\begin{definition}~\label{def:OPE}
\noindent \textbf{Order Preserving Encryption~\mbox{\boldmath$(\ope)$}:} is a deterministic symmetric encryption scheme whose encryption preserves the numerical ordering of the plaintexts, i.e. for any two messages $m_1$ and $m_2 \:~s.t.~\: m_1 \leq m_2$, we have $c_1\as\OEnc{K}{m_1}$ $\leq c_2\as\OEnc{K}{m_2}$ \cite{boldyreva2009order}, with $c\as\OEnc{K}{m}$ is order preserving encryption of a message $m\in\{0,1\}^{d}$ under key $K$, where $d$ is the block size of \ope.
\end{definition}

Note that communications are made over a secure (authenticated) channel maintained with a symmetric key (e.g., via SSL/TLS as in Algorithm \ref{alg2}) to ensure confidentiality and authentication. For the sake of brevity, we will only write encryptions but not the authentication tags (e.g., Message Authentication Codes~\cite{JonathanKatzModernCrytoBook}) for the rest of the paper.

%
%

\section{\LPGW}
\label{sec:lpgw}
We now present our proposed scheme that we call \LPGW~(location privacy for 3-party spectrum sensing architecture), which offers high location privacy and low overhead, and uses an additional entity in the network, referred to as Gateway (\gw) (thus "3P" refers to the 3 parties: \su s, \fc, and \gw). \gw~enables a higher privacy by preventing \fc~from even learning the order of encrypted \rss~values of \su s which was allowed in \LPOS~\cite{grissa2015location}. \gw~also learns nothing but secure comparison outcome of \rss~values and \ta, as in \ym~but only using \ope. Thus, no entity learns any information on \rss~or~\ta~beyond a pairwise secure comparison, which is the minimum information required for a voting-based decision.

$\bullet$ {\em Intuition}: The main idea behind \LPGW~is simple yet very powerful: We enable \gw~to privately compare \nbr~distinct \ope~encryptions of \ta~and \rss~values, which were computed under \nbr~pairwise keys established between \fc~and \su s. These \ope~encrypted pairs permit \gw~to learn the comparison outcomes without deducing any other information. \gw~then sends these comparison results to \fc~to make the final decision. \fc~learns no information on \rss~values and \su s cannot obtain the value of \ta, which complies with our Security Objectives \ref{obj:SecurityObj-1} and \ref{obj:SecurityObj-2}. Note that \LPGW~relies {\em only on symmetric cryptography} to guarantee the location privacy of \su s. Hence, it is the {\em most computationally efficient and compact} scheme among all alternatives (see Section \ref{sec:PerformanceAnalysis}),  but with an additional entity in the system.

\LPGW~is described in Algorithm~\ref{alg2} and  outlined below. 

 \begin{algorithm}[h!]
\caption{\LPGW~Algorithm}\label{alg2}
\begin{algorithmic}[1]
\Statex   \textbf{Initialization}: Executed only once.
\State \fc~sets energy sensing, optimal voting thresholds \ta, \lam, and weights vector $\boldsymbol{w} \gets \boldsymbol{1}$, respectively.
\State Entities establish private pairwise keys and maintain authenticated secure channels (e.g., via SSL/TLS) as follows:
\Statex  $\bullet$~$\key_{\fc,i}$ between \fc~and each user $\user_i$, $i=1,\ldots ,\nbr$.
\Statex  $\bullet$~$\key_{\gw,i}$ between \gw~and each user $\user_i$, $i=1,\ldots ,\nbr$.
\Statex  $\bullet$~$\key_{\fc,\gw}$ between \fc~and \gw.
\State \fc~computes $\ciphFC_{i} \gets \Enc{\key_{\fc,\gw}}{\OEnc{\key_{\fc,i}}{\ta}}$, $i=1,\ldots ,\nbr$ and sends $\{\ciphFC_i\}_{i=1}^{n}$ to \gw. \label{alg1:enc1}
\hspace{20pt}\algrule
\Statex   \textbf{Private Sensing}: Executed every sensing period $t_w$
\State $\user_i$ computes $\ciphU_{i} \gets \Enc{\key_{\gw,i}}{\OEnc{\key_{\fc,i}}{\rss_i}}$, $i = 1,\ldots,\nbr$ and sends $\{\ciphU_i\}_{i=1}^{n}$ to \gw.\label{alg1:enc2}
\State \gw~obtains $\OEnc{\key_{\fc,i}}{\ta} \gets \Dec{\key_{\fc,\gw}}{\ciphFC_{i}}$ and   $\OEnc{\key_{\fc,i}}{\rss_i} \gets \Dec{\key_{\gw,i}}{\ciphU_{i}}$, $i = 1,\ldots,\nbr$.
\For{\texttt{$i = 1,\ldots,\nbr$}}
\If {$\OEnc{\key_{\fc,i}}{\rss_i} < \OEnc{\key_{\fc,i}}{\ta}$} $\bin_i \gets 0$\label{alg1:comp}
\Else \:$\bin_i \gets 1$
\EndIf
\EndFor
\State \gw~computes $\ciphGW \gets \Enc{\key_{\fc,\gw}}{\{\bin_i\}_{i=1}^{n}}$ and sends $\ciphGW$ to \fc. \label{alg1:enc}

\State \fc~decrypts $\ciphGW$ and computes $\votes\as \sum\limits_{i=1}^{\nbr}w_i\times b_i$ \label{alg1:sum}

\If {$\votes \geq  \lam$}  $\decision$ $ \gets $ Channel busy
\Else \: $\decision$ $ \gets $ Channel free
\EndIf
\State \fc~updates the credibility score $\varphi_i$ and weight $w_i$ of each user $U_i$ as in equations~\ref{cred} and \ref{weight} for $i=1,\ldots,\nbr$ \label{upWeight}

\Return $\decision$
\hspace{20pt}\algrule
\Statex   \textbf{Update after $\mathcal{G}$ Membership Changes or Breakdown}:
\State If a user joins the network, it needs to establish a pairwise secret key with \fc~and \gw. If \su(s) join/leave or breakdown, \lam~is updated as \lam'.
\State Follow the private sensing steps with new \lam'.
\end{algorithmic}
\end{algorithm}

 $\bullet$~{\em Initialization:} Let $(\En,\De)$ be IND-CPA secure~\cite{JonathanKatzModernCrytoBook} block cipher (e.g. \aes) encryption/decryption operations. \fc~establishes a secret key with each \su~and \gw. \gw~establishes a secret key with each \su. \fc~encrypts \ta~with \ope~using $\key_{\fc,i}$, $i=1 \ldots \nbr$. \fc~then encrypts \ope~ciphertexts with \En~using $\key_{\fc,\gw}$ and sends these $\ciphFC_{i}$s to \gw, $i=1 \ldots \nbr$. Since these encryptions are done offline at the beginning of the protocol, they do not impact the online private sensing phase. \fc~may also pre-compute a few extra encrypted values in the case of new users joining the sensing.

 $\bullet$~{\em Private Sensing:} Each $\user_i$~encrypts $\rss_i$ with \ope~using $\key_{\fc,i}$, which was used by \fc~to \ope~encrypt \ta~value. $\user_i$ then encrypts this ciphertext with \En~using key $\key_{\gw,i}$, and sends the final ciphertext $\ciphU_{i}$ to \gw. \gw~decrypts $2\nbr$ ciphertexts $\ciphFC_{i}$s and $\ciphU_{i}$s with \De~using $\key_{\fc,\gw}$ and $\key_{\gw,i}$, which yields \ope~encrypted values. \gw~then compares each \ope~encryption of \rss~with its corresponding \ope~encryption of \ta. Since both were encrypted with the same key, \gw~can compare them and conclude which one is greater as in Step~\ref{alg1:comp}. \gw~stores the outcome of each comparison in a binary vector $\mbox{\boldmath$ \bin$}$, encrpyts and sends it to \fc. Finally, \fc~compares the summation of votes \votes~to the optimal voting threshold \lam~to make the final decision about spectrum availability and updates the reputation scores of the users.

  $\bullet$~{\em Update after $\mathcal{G}$ Membership Changes or Breakdown:} Each new user joining the sensing just establishes a pairwise secret key with \fc~and \gw. This has no impact on existing users. If some users leave the network, \fc~and \gw~remove their secret keys, which also has no impact on existing users. In both cases, and also in the case of a breakdown or failure, \lam~must be updated accordingly.

\section{Security Analysis}
\label{sec:SecAnalysis}

\begin{table*}[t!]

\centering  \caption{Computational overhead comparison} \label{tab:Table2}

\resizebox{\textwidth}{!}{%
\renewcommand{\arraystretch}{1.25}{
\begin{tabular}{||c||c|c|c|c||}

\hline \multicolumn{1}{||c||}{\multirow{2}{*}{\textbf{\em Scheme}}}  & \multicolumn{4}{|c||}{\textbf{Computation}} \\ \cline{2-5}

\multicolumn{1}{||c||}{} &  \textbf{ {\em FC}} & \multicolumn{2}{c|}{\textbf{ {\em SU}}} & \textbf{ {\em GW}}\\ \hline
 \hline   \multicolumn{1}{||c||}{\textbf{\LPGW}}  &  $\De + \chr\cdot(\En+\ope_E)$ & \multicolumn{2}{c|}{$\ope_E + \En$ }& $\nbr \cdot\De + \En$  \\ 
\hline  \multicolumn{1}{||c||}{\textbf{\LPOS}}  &  $1/2 \cdot(2+log\:\nbr)\cdot\gam \cdot|p|\cdot Mulp$ & \multicolumn{2}{c|}{$(2\gam\cdot |p|+2\gam)\cdot Mulp+\ope +2\sr\cdot log\: \nbr \cdot PMulQ$} & - \\
 \hline \multicolumn{1}{||c||}{\PPSS} &  $H + (\nbr+2) \cdot Mulp + (2^{\gam-1}\cdot\nbr + 2) \cdot Expp$ & \multicolumn{2}{c|}{$H + 2Expp + Mulp$} & -\\
\hline \multicolumn{1}{||c||}{\PDAFT} & $2Exp\rsa^2 + Inv\rsa^2 + \srv \cdot Mul\rsa^2$ &\multicolumn{2}{c|}{$ 2Exp\rsa^2 + Mul\rsa^2$ } & $\nbr\cdot Mul\rsa^2$\\ \hline
\end{tabular}}}

\flushleft{\scriptsize{ \textbf{(i) Variables:} $\kap$ security parameter, $\rsa$: modulus in Paillier, $p$: modulus of El Gamal, $H$: cryptographic hash operation, $K$: secret group key of \ope. $Expu$ and $Mulu$ denote a modular exponentiation and a modular multiplication over modulus $u$ respectively, where $u \in \{\rsa, \rsa^2, p\}$. $Inv\rsa^2$: modular inversion over $\rsa^2$, $PMulQ$: point multiplication of order $Q$, $PAddQ$: point addition of order $Q$. $\srv$: number of servers needed for decryption in \PDAFT.
\textbf{(ii)  Parameters size:} For a security parameter $\kappa = 80$, suggested parameter sizes by {\em NIST 2012} are given by : $|\rsa| = 1024$, $|p| = 1024$, $|Q|=192$ as indicated in \cite{keylength}. \textbf{(iii) OPE}: we rely on \ope~scheme proposed by Boldyreva~\cite{boldyreva2009order} for our evaluation because of its popularity and public implementation but our schemes can use {\em {any secure}} \ope~scheme (e.g.,~\cite{boldyreva2009order,popa2013ideal,kerschbaum2014optimal}) as a building block. \textbf{(v) \En}: We rely on \aes~\cite{daemen1999aes}\footnote{AES is a symmetric block cipher adopted by the U.S. government and known to be the strongest symmetric crypto algorithm.} as our (\En,\De) for our cost analysis.
}}

\vspace{-3mm}

\end{table*}

We first describe the underlying security primitives, on which our schemes rely, and then precisely quantify the information leakage of our schemes, which we prove to achieve our Security Objectives \ref{obj:SecurityObj-1} and \ref{obj:SecurityObj-2}.

\begin{fact}\label{fact:IdealSecOPE}
 An \ope~is \emph{indistinguishable under ordered chosen-plaintext attack (IND-OCPA)} \cite{boldyreva2009order} if it has no leakage, except the order of ciphertexts (e.g. \cite{popa2013ideal,kerschbaum2014optimal}).
\end{fact}

Let \En~and \OEncc~be {\em IND-CPA secure}~\cite{JonathanKatzModernCrytoBook} and {\em IND-OCPA secure} symmetric ciphers, respectively. $(\{\rss_{i}^{j}\}_{i=1,j=1}^{n,l},\tau)$ are \rss~values and \ta~of each $\user_i$ and \fc~for sensing periods $j=1,\ldots,l$ in a group $\mathcal{G}$. $(\LU,\LF,\LG)$ are history lists, which include all values learned by entities $\user_i$, \fc~and \gw, respectively, during the execution of the protocol for all sensing periods and membership status of $\mathcal{G}$. Vector \V~is a list of IND-CPA secure values transmitted over secure (authenticated) channels. \V~may be publicly observed by all entities including external attacker \A. Hence, \V~is a part of all lists $(\LU,\LF,\LG)$. Values (jointly) generated by an entity such as cryptographic keys or variables stored only by the entity itself (e.g., \lam, \pr) are not included in history lists for brevity.

\begin{mytheorem} \label{the:Security:3P}
Under Security Assumptions \ref{asm:asm1} and \ref{asm:asm2}, \LPGW~leaks no information on $(\{\rss_{i}^{j}\}_{i=1,j=1}^{n,l},\tau)$ beyond IND-CPA secure $\{\V^{j}\}_{j=1}^{l}$, IND-OCPA secure pairwise order  $\{\OEnc{\key_{\fc,i}}{\rss_{i}^{j}},$ $\OEnc{\key_{\fc,i}}{\ta}\}_{i=1,j=1}^{n,l}$ to \gw~and $\{b_{i}^{j}\}_{i=1,j=1}^{n,l}$ to \fc.
\end{mytheorem}
\noindent {\em Proof:} $\V^{j}=\{c_{i}^{j},\ciphU_{i}^{j},\ciphGW^{j}\}_{i=1,j=1}^{n,l}$, where $\{c_{i}^{j}\}_{i=1,j=1}^{n,l}$ and $\{\ciphU_{i}^{j},\ciphGW^{j}\}_{i=1,j=1}^{n,l}$ are generated at the initialization and privacy sensing in Algorithm \ref{alg2}, respectively. History lists are as follows for each sensing period $j=1,\ldots,l$:
\begin{eqnarray*} \label{eq:HistorList3P}
\LU & = & \V^{j},~~~\LF = (\{b_{i}^{j}\}_{i=1,j=1}^{n,l},\V^{j}), \\
\LG & = & (\{\OEnc{\key_{\fc,i}}{\rss_{i}^{j}},\OEnc{\key_{\fc,i}}{\ta}\}_{i=1,j=1}^{n,l},\V^{j},\\& &\{b_{i}^{j}\}_{i=1,j=1}^{n,l})
\end{eqnarray*}
Variables in $(\LU,\LF,\LG)$ are IND-CPA secure and IND-OCPA secure, and therefore leak no information beyond the pairwise order of ciphertexts to \gw~by Fact \ref{fact:IdealSecOPE}.

Any membership status update on $\mathcal{G}$ requires an authenticated channel establishment or removal for joining or leaving members, whose private keys are independent from each other. Hence, history lists $(\LU,\LF,\LG)$ are computed identically as described above for the new membership status of $\mathcal{G}$, which are IND-CPA secure and IND-OCPA secure.\hfill$\square$

\begin{mycorollary} \label{Cor:SecurityObjectives}
Theorem \ref{the:Security:3P} guarantees that in our scheme, RSS values and \ta~are IND-OCPA secure for all sensing periods and membership changes. Hence, our scheme achieves Objectives \ref{obj:SecurityObj-1} and \ref{obj:SecurityObj-2} as required.
\end{mycorollary}

\section{Performance Evaluation}
\label{sec:PerformanceAnalysis}
We now evaluate our proposed scheme, \LPGW, by comparing it to existent approaches that we briefly explain below.

\subsection{Existing Approaches}

\PPSS~\cite{li2012location} uses secret sharing and the Privacy Preserving Aggregation (PPA) process proposed in~\cite{shi2011privacy} to hide the content of specific sensing reports and uses dummy report injections to cope with the DLP attack.
\LPOS~\cite{grissa2015location} also uses \ope~but in a completely different way than how we use it in this paper. Users \ope~encrypt their \rss~values, send them to \fc~which, based on the order of the encrypted \rss s, performs at worst a logarithmic number of yao's millionaires secure comparisons \cite{lin2005efficient} between \ta~and \rss s and then makes a final decision about spectrum availability.
\PDAFT~\cite{chen2014pdaft} combines Paillier cryptosystem~\cite{paillier1999public} with Shamir's secret sharing~\cite{shamir1979share}, where a set of smart meters sense the consumption of different households, encrypt their reports using Paillier, then send them to a gateway. The gateway multiplies these reports and forwards the result to the control center, which selects a number of servers (among all servers) to cooperate in order to decrypt the aggregated result. \PDAFT~requires a dedicated gateway, just like \LPGW, to collect the encrypted data, and a minimum number of working servers in the control center to decrypt the aggregated result.

\subsection{Performance Analysis and Comparison}
We focus on communication and computational overheads. We consider the overhead incurred during the sensing operations but not that related to system initialization (e.g. key establishment), where most of the computation and communication is done offline. We model the membership change events in the network as a random process \ran~that takes on 0 and 1, and whose average is $\sr$. $\ran=0$ means that no change occurred in the network and $\ran=1$ means that some users left/joined the sensing task. Let \chr(t)~be a function that models the average number of users that join the sensing at the current sensing period $t$, where
\begin{equation*}
\chr(t) = \begin{cases} \nbr(t) - \nbr(t-1), & \mbox{if } \nbr(t) - \nbr(t-1) > 0 \; \& \; \ran(t)=1\\ 0, & \mbox{otherwise }\end{cases}
\end{equation*}

The execution times of the different primitives and protocols were measured on a laptop running Ubuntu 14.10 with 8GB of RAM and a core M 1.3 GHz Intel processor, with cryptographic libraries MIRACL~\cite{miracl}, Crypto++~\cite{crypto++} and {\em Louismullie}'s Ruby implementation of \ope~\cite{opeRuby}.

\noindent \textbf{Computational Overhead}: Table \ref{tab:Table2} provides an analytical computational overhead comparison including the details of variables, parameters and the overhead of building blocks.

 In \LPGW, \fc~requires only a small constant number of $(\De,\En,\ope)$ operations. An \su~requires one \ope~and \En~encryptions of its \rss. Finally, \gw~requires one \De~operation per user and one \En~of vector $\mbox{\boldmath$\bin$}$. All computations in \LPGW~rely on only symmetric cryptography, which makes it {\em the most computationally efficient scheme among all alternatives}.

\begin{figure}[t!]
  \centering
  \subfigure[Computational overhead]{\includegraphics[scale=0.2]{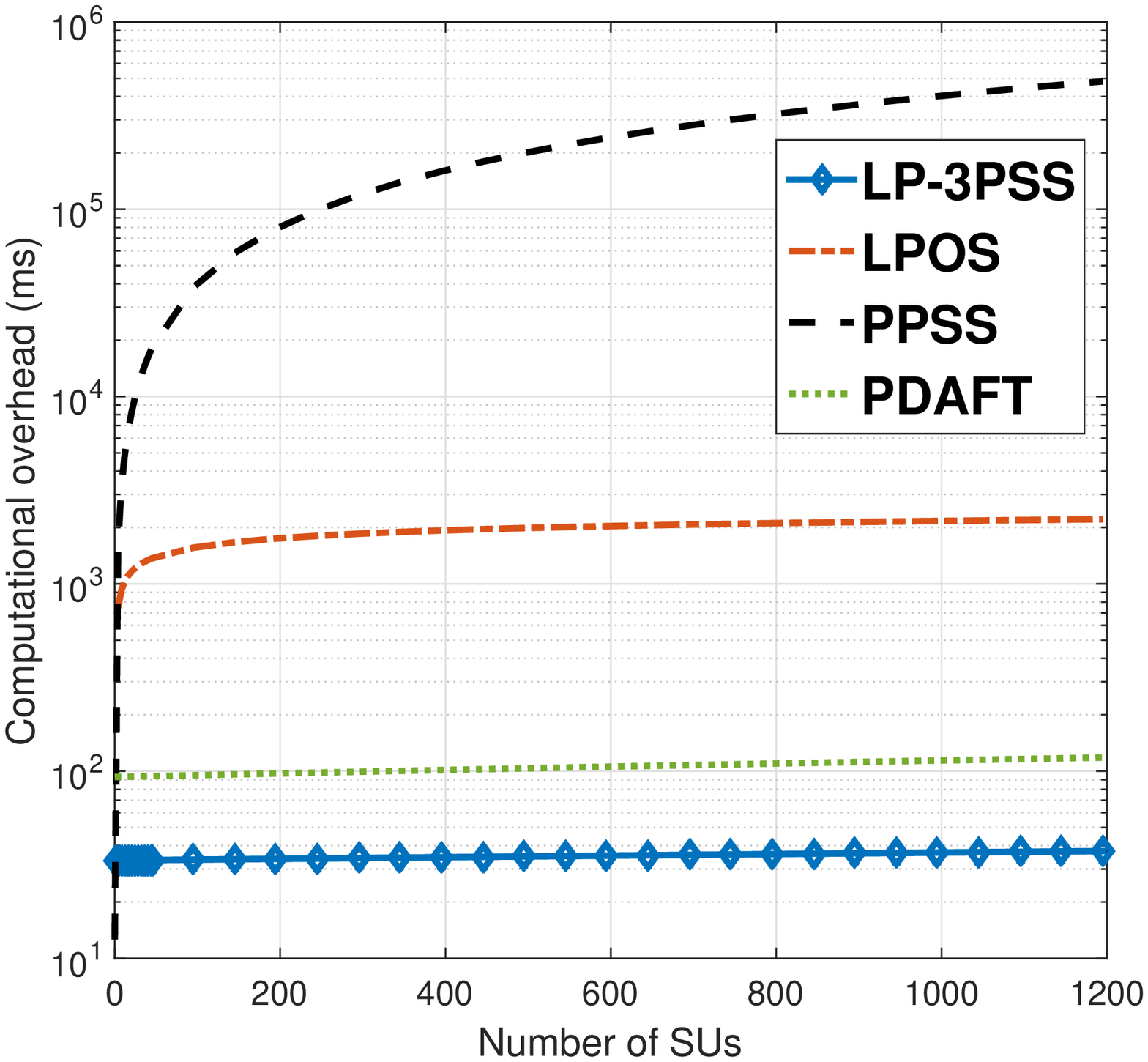}\label{fig:comp_overhead}}
  \subfigure[Communication overhead]{\includegraphics[scale=0.2]{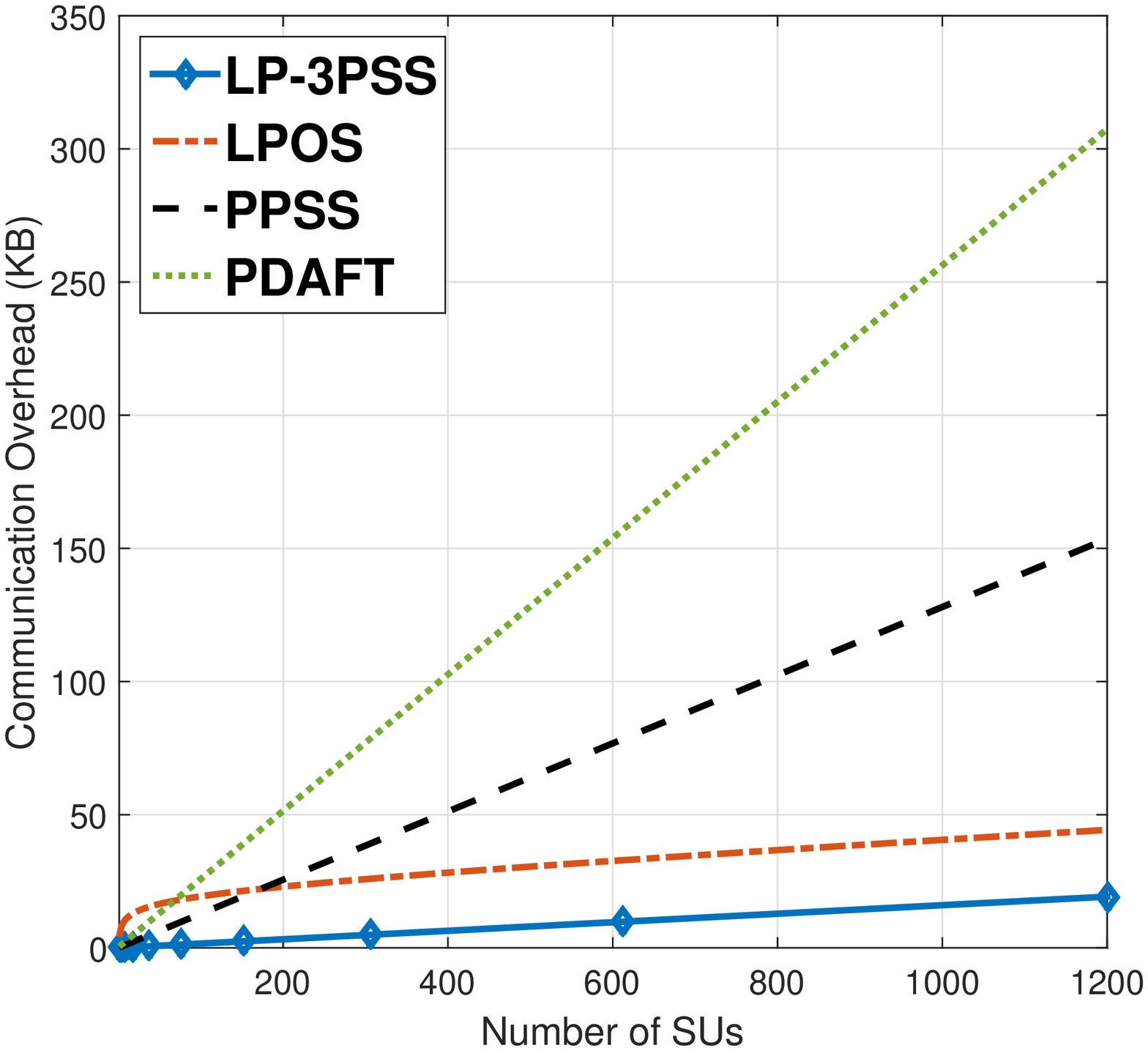}\label{fig:comm_overhead}}
  \caption{Performance compariosn, $\kap =80$, $\chr = 5$, $\sr = 20\%$} \label{fig:perfComm}
\end{figure}

For illustration purpose, we plot in Fig.~\ref{fig:comp_overhead} the system end-to-end computational overhead of the different schemes. Fig.~\ref{fig:comp_overhead} shows that \LPGW~is several order of magnitudes faster than the other schemes including \LPOS, that we proposed in a previous work, for any number of users.

 \begin{table}[t!]

\centering  \caption{Communication overhead comparison} \label{tab:Table3}
\renewcommand{\arraystretch}{1.25}{
\begin{tabular}{||c||c||}

\hline \multicolumn{1}{||c||}{\textbf{\em Scheme}}   &  \multicolumn{1}{|c||}{\textbf{Communication}} \\

\hline \hline \multicolumn{1}{||c||}{\textbf{\LPGW}} & $ (\nbr + 1 )\cdot \blck$ \\
\hline  \multicolumn{1}{||c||}{\textbf{\LPOS}} & $ 2\gam \cdot |p| \cdot (2+log\:\nbr)+ \nbr \cdot \epsilon_{\ope}+\sr\cdot|Q| \cdot log\: \nbr$  \\
\hline \multicolumn{1}{||c||}{\PPSS} & $ |p|\cdot \nbr + \chr\cdot\sr\cdot |p| \cdot \nbr$  \\
\hline \multicolumn{1}{||c||}{\PDAFT} & $|\rsa|\cdot (2(\nbr+1)+\chr)$  \\ \hline

\end{tabular}}
\flushleft{\scriptsize{
$\epsilon_{\ope} = 128\:bits$: maximum ciphertext size obtained under \ope~encryption, \blck: size of ciphertext under \En.
}}
\vspace{-3mm}
\end{table}

\noindent \textbf{Communication Overhead}:
Table~\ref{tab:Table3} provides the analytical communication overhead comparison. \LPGW~requires (\nbr+1) \En~ciphertexts and single \ciphGW, which are significantly smaller than the ciphertexts transmitted in the other schemes.

We further compare our scheme with its counterparts in terms of communication overhead in Fig.~\ref{fig:comm_overhead}. Fig.~\ref{fig:comm_overhead} shows that \LPGW~has the smallest communication overhead since, again, it relies on symmetric cryptography only.  \PPSS~and \PDAFT~have a very high communication overhead due to the use of expensive public key encryptions (e.g., Pailler~\cite{paillier1999public}).

Overall, our performance analysis indicates that \LPGW~is significantly more efficient than all other counterpart schemes in terms of computation and communication overhead, even for increased values of the security parameters, but with the cost of including an additional entity.


\section{Conclusion}
\label{sec:Conclusion}
We developed an efficient scheme for cooperative spectrum sensing that protects the location privacy of \su s with a low cryptographic overhead while guaranteeing an efficient spectrum sensing. Our scheme is secure and robust against users dynamism, failures, and user maliciousness. Our performance analysis indicates that our scheme outperforms existing alternatives in various metrics.

\section*{Acknowledgment}
This work was supported in part by the US National Science Foundation under NSF award CNS-1162296.

\small{
\bibliographystyle{IEEEtran}
\bibliography{IEEEabrv,./references}
}

\end{document}